\documentclass[letter]{jpconf}
\usepackage{graphicx}
\usepackage{amsmath}
\usepackage{epstopdf}

\newcommand{\beqn}{\begin{equation}}
\newcommand{\eeqn}{\end{equation}}
\newcommand{\req}[1]{Eq.\,(\ref{#1})}
\newcommand{\GeV}{\text{ GeV}}
\newcommand{\MeV}{\text{ MeV}}
\newcommand{\keV}{\text{ keV}}
\newcommand{\eV}{\text{ eV}}
\newcommand{\meV}{\text{ meV}}

\begin{document}

\title{Traveling Through  the Universe:\\ Back in Time  to the  Quark-Gluon Plasma Era}

\author{Johann Rafelski$^1$,  and Jeremey Birrell$^2$}

\address{$^1$Department of Physics, and $^2$Program in Applied Mathematics\\
The University of Arizona, Tucson, AZ 85721, USA}
 

\begin{abstract}
We   survey  the early history of the discovery of quark gluon plasma and the early history of the Universe, beginning with  the present day and reaching  deep into QGP and almost  beyond. We introduce  cosmological Universe dynamics and connect the different Universe epochs with one another. We  describe some of the many remaining open questions that emerge.
\end{abstract}

\vskip -10.cm 
{\small \centerline{BASED ON INVITED LECTURES PRESENTED  AT:}
\centerline{\phantom{and}}
\centerline{STRANGENESS IN QUARK MATTER MEETING, BIRMINGHAM, UK, July  2013}
\centerline{and}
\centerline{APS DIVISION OF NUCLEAR PHYSICS, Newport News, October  2013}
}
\vskip 8.3 cm

\section{A long, long time ago in our Universe}
Matter emerged from  a primordial state   we call quark-gluon plasma (QGP). We describe how relativistic heavy ion collision laboratory experiments shape how we think about the origin of matter, and we study the quark-hadron Universe applying these results. The discovery of  QGP  can be dated back to the CERN press event of February 10, 2000, when the evidence for ``a new state of matter discovered at CERN'' was presented. These experimental results originated in the relatively low SPS energy scale.  

`QGP' is not a new particle but a paradigm-shift of how we understand matter in extreme conditions. Thus there is no clear criterion by which one can claim an experimental  discovery. For this reason QGP was rediscovered  again with new experimental results obtained at the order of magnitude higher RHIC collision energies. In addition to  confirming the CERN results, RHIC produced new puzzling phenomena; some will be discussed below. The  circumstance repeats for the third time today: LHC data confirms SPS and RHIC results, and is offering another rich field of new experimental results.  

Since  no one plans to announce the QGP discovery at the LHC, we conclude that QGP has gained  considerable acceptance as a new form of matter. This then suggests using this  new paradigm in the study of other phenomena. Here we will  travel  back in time   to explore the properties of the Universe when it was much hotter and much denser, journeying beyond the era of hadrons into the quark Universe.  We will briefly visit three stages in the Universe evolution, beginning from today's condition as we go back in time:
\begin{enumerate}
\item  The period reaching back to shortly after big-bang nucleosynthesis (BBN) completed  is the directly `observable' history of the Universe, in the sense that some aspects of this evolution are directly visible today and are being analyzed in depth. Our knowledge is based primarily on two observables. The cosmic microwave background (CMB), the  red-shifted heat of the big-bang, reaches us without much interaction from an epoch which can be dated back to about 1/1000 of the age of the Universe. The cosmic abundance of light elements is interpreted as the consequence of nuclear burning in the early Universe that ended when the Universe  was a fraction of an hour old.
\item  The era connecting the period just after baryon antimatter disappeared to BBN passes through 7 orders of magnitude in time, from 1ms to 1000 s. The Universe emerging from the baryon antimatter  annihilation phase still comprises  muons and pions. When these disappear  the Universe expands,  pushed apart by effectively massless $e^+e^-\gamma$-QED plasma, and  neutrinos. The three neutrino flavors decouple just before  BBN and become a separate radiation fraction, while   $e^+e^-$-annihilation continues through the BBN process. There is good hope of identifying observables from this period which could still be connected to the present day Universe. Current interest is centered on improving the understanding of the  energy content and momentum distribution of neutrino cosmic background radiation. 
\item The earliest stage we can now begin to discuss in detail is the  evolution of the Universe from about 10ps to the end of QGP hadronization at 20$\mu$s and the end of baryon antimatter   annihilation at 1 ms, thus 8 orders of magnitudes in time scale. During this period we see the Universe emerge from the electro-weak symmetry restored stage into the quark stage, and evolve into a dense hadron stage, which in turn disappears when both antibaryons annihilate and mesons decay. It is in this period that our recent experimental heavy ion insights into the quark-hadron Universe become of use. 
\end{enumerate}
Because the QGP hadronization process is complex and could involve inhomogeneity of the matter distribution, there has always been the intuitive feeling  that some macroscopic observables will ultimately emerge that relate to matter distribution in the Universe. Here we recall  the mystery of the large scale structure of visible matter, which is localized  to the surface or even to the edges of bubble-like boundaries with a scale difference of  matter structure thickness size to the matter void size being as large as 50. 

After presenting in section \ref{QGPdisc} a few motivational remarks about the process of discovery and importance of QGP, we offer in section \ref{cosmo} a short introduction to the essential elements of cosmology  required to follow our trip back in time. In sections \ref{recomb}, \ref{nudecoup}, \ref{QGPera}  we offer in turn discussions of these stages of the Universe evolution. We connect the three eras and describe some of the implications of our findings in section \ref{Eralink}. We complete our discussion with a short listing of open questions in section \ref{conclude}. The results and discussion presented here complement our recent related considerations which were addressing specific  questions about  the QGP era in the Universe~\cite{Fromerth:2012fe,Rafelski:2013qeu}. 

\section{How we got to QGP from the hadron side}\label{QGPdisc}
The 20 year long march to the QGP announcement at CERN  also started at CERN. In the early days this was the one and only location in the world with a considerable tradition of thermal particle physics. One of us (JR) arrived at CERN in 1977 when Hagedorn temperature $T_H=160$ MeV was already a local  trademark, introduced to characterize multiparticle production in what were at the time high energy $pp$ collisions. In search of an explanation of the cause for  this thermal  behavior, a rather unexpected novel feature of  particle physics, Rolf Hagedorn proposed that point hadrons could be made of other hadrons, within a `bootstrap' concept. Once  quarks as building blocks of hadrons were introduced, the bootstrap model in which hadrons are made of hadrons was justified  in terms of small domains in space filled with quarks. However,  the quark-hadron bootstrap was theoretically  consistent only once  hadrons were allowed to have a proper volume. This refined model allowed the study of how hadrons, occupying a finite volume and stressed in collision, dissolve  into their constituents, creating in this way a hot blob of quarks we call QGP today. At the 1984 Quark Matter meeting in Helsinki, Hagedorn described  these developments in the last pages of the very accurate  historical reminiscences  ``How we got to  QCD  matter from the hadron side by trial and error,''~\cite{Hagedorn:1984hz}.
 
With Hagedorn reporting the tale of the theoretical formulation of QGP formation in heavy ion collisions nearly 30 years ago, and strangeness as a signature of QGP  already a well developed idea at that time~\cite{Rafelski:1982ii}, what exactly have  we done since? Many ideas and developments were devoted  to alternate  diagnostic tools, and many related novel theoretical ideas emerged. The situation is presented succinctly within the abstract of a recent review of Jacak and Muller~\cite{Jacak:2012dx} ``\ldots this new state of matter is a nearly ideal, highly opaque liquid. A description based on string theory and black holes in five dimensions has made the quark-gluon plasma an archetypal strongly coupled quantum system.'' 

These courageous words relate to the finding that QGP matter  flows  with minimal friction. This is somewhat surprising since we also observe both at RHIC and prominently at LHC, that in  QGP  high energy parton jets rapidly quench. These two observations in fact are a contradiction, since QCD naively predicts just the opposite,  specifically that at the low energy scales  the QCD interactions should be hindering flow, and at the high energy scale of jets,  the parton interaction is weakening, removing quenching. Clearly we do have some remaining theoretical challenges in the study of the behavior and properties of QGP -- we note that  no one is able to precisely tell based on QCD when the expanding QGP   `hadronizes',  breaking up into individual hadrons, and conversely, when as function of heavy ion collision energy we form QGP.  

An important epistemological question is how the introduction of QGP into the physics vocabulary  changed our global understanding of natural phenomena:\\
a) QGP exhibits quark deconfinement.\\  
b) Conversely, the  quark confinement effect makes hadrons massive, explaining the origin of 99\% of the mass of matter~\cite{Kronfeld:2012ym}.\\ 
c) Quark confinement is recognized as a transport property of empty space, the structured quantum vacuum, rather than a direct property of the quark-quark interaction. \\
d)  The dominant interaction of quarks (and gluons) is with empty space, the quantum vacuum, a major departure from two body interactions being the dominant force in elementary processes.

This insight creates another paradigm shift; the vacuum state is strongly structured. With this, a further challenge arises relating to the understanding of quantum vacuum structures both in principle and at a technical level. The technical effort has been centered for many years on the ever improving  lattice-QCD description of the deconfinement phenomenon~\cite{Philipsen:2012nu}.  However, we lack a true in principle understanding of how the quantum vacuum structure acts. What is the mechanism that we could present in the classroom to explain why the mass of a small quark bubble such as a proton is at the scale of 1 GeV and not 0.5 or 2 GeV?  Where  exactly does the scale come from? 

\section{Standard Cosmology}\label{cosmo}
In order to travel back in time to the period when QGP dominated the Universe, we first need to  elaborate on the relation between the expansion dynamics of the Universe and temperature.  For this purpose we need some preparation in the standard  cosmological (FLRW-Universe) model. We use the spacetime metric
\beqn\label{metric}
ds^2=c^2dt^2-a^2(t)\left[ \frac{dr^2}{1-kr^2}+r^2(d\theta^2+\sin^2(\theta)d\phi^2)\right]
\eeqn
characterized  by the scale parameter $a(t)$  of a homogeneous spatial Universe. The geometric parameter $k$ identifies the geometry of the spacial hypersurfaces defined by comoving observers. Space is a flat-sheet for the observationally preferred value $k=0$ \cite{Planck}. The global Universe dynamics can be characterized by two  quantities: the Hubble parameter  $H$, a strongly time dependent quantity on cosmological time scales,  and the deceleration parameter $q$:
\beqn\label{dynamic}
\frac{\dot a }{a}\equiv H(t) ,\quad \frac{\ddot a}{a}=-qH^2,\quad 
q\equiv -\frac{a\ddot a}{\dot a^2},\quad \dot H=-H^2(1+q). 
\eeqn

The Einstein equations are:
\beqn\label{Einstine}
G^{\mu\nu}=R^{\mu\nu}-\left(\frac R 2 +\Lambda\right) g^{\mu\nu}=8\pi G_N T^{\mu\nu},  
\quad R= g_{\mu\nu}R^{\mu\nu}, \quad T^\mu_\nu =\mathrm{diag}(\varepsilon, -P, -P, -P).
\eeqn
It is common to absorb the Einstein cosmological constant $\Lambda$ into the energy and pressure
\beqn\label{EpsLam}
\varepsilon_\Lambda=\frac{\Lambda}{8\pi G_N}, \qquad P_\Lambda=-\frac{\Lambda}{8\pi G_N}
\eeqn
and we implicitly consider this done from now on. 

One should note that the bag constant $ {\cal B} $ of the quark-bag model has the same behavior in regard to energy and momentum as has the Einstein cosmological parameter $ {\cal B} \leftrightarrow \Lambda/8\pi G_N$.    $ {\cal B} $ adds  positively to the energy density but negatively to the pressure, counteracting the positive particle pressure. Contrary to the initial expectation based on the quark-bag model where quark pressure is in equilibrium with the bag constant, in the dynamical Universe the appearance of such a bag term will accelerate the expansion just like today where we see an acceleration due to dark energy. The parallel meaning of $ {\cal B}$ and $ \Lambda/8\pi G_N$ relies on both quantities acting within the volume of their respective  `Universe', in the sense that $ {\cal B}$ is strictly and only present within the volume of quark blobs -- hadrons, or QGP.

Two dynamically independent equations arise using the metric \req{metric} in \req{Einstine}:
\beqn\label{hubble}
\frac{8\pi G_N}{3} \varepsilon =  \frac{\dot a^2+k}{a^2}
=H^2\left( 1+\frac { k }{\dot a^2}\right),
\qquad
\frac{4\pi G_N}{3} (\varepsilon+3P)  =-\frac{\ddot a}{a}=qH^2.
\eeqn
We can eliminate the strength of the interaction, $G_N$,  solving both these equations for ${8\pi G_N}/{3}$, and equating the result to find a relatively simple constraint for the deceleration parameter:
\beqn\label{qparam}
q=\frac 1 2 \left(1+3\frac{P}{\varepsilon}\right)\left(1+\frac{k}{\dot a^2}\right).
\eeqn
For $k=0$ note  that in a  matter-dominated Universe where $P/\varepsilon<<1$ we have $q\simeq 1/2$; for a radiative Universe where $3P=\varepsilon$ we find $q\simeq 1 $; and  in a dark energy Universe in which $P=-\varepsilon$  we find $q=-1$. We will work here within the   standard cosmology model with $k=0$.

As must be the case for any solution of Einstein's equations,   \req{hubble} implies that the energy momentum tensor of matter is divergence free:
\beqn\label{divTmn}
T^{\mu\nu}||_\nu =0 \Rightarrow -\frac{\dot\varepsilon}{\varepsilon+P}=3\frac{\dot a}{a}=3H.
\eeqn
The same relation also follows from  conservation of entropy, $dE+PdV=TdS=0,\  dE=d(\varepsilon V),\  dV=d(a^3)$. A dynamical evolution equation for $\varepsilon(t)$ arises once we combine \req{divTmn} with \req{hubble},  eliminating $H$.   Given an equation of state $P(\epsilon)$, solution of this equation describes the dynamical evolution of matter in the Universe. We can study the entropy conserving Universe evolution  both forwards and backwards in time. However, there are entropy producing processes such as antibaryon annihilation which will need further attention in the future.

\section{Back in time to BBN}\label{recomb}
Given  equations of state connecting $\varepsilon$ and $P$ for the various components of the Universe matter content,  \req{hubble} and \req{divTmn} for $k=0$ are immediately soluble and the outcome is widely discussed for the period of time spanning BBN to the present era,  passing through photon freeze-out at the ion recombination era. In the following we use the mix of matter  (31\%) and dark energy (69\%) with photon and neutrino backgrounds favored by the latest Planck results \cite{Planck}, where we gave two neutrino species mass of $m_\nu=30\meV$ and a third neutrino remains  massless -- other mass choices are possible within present day constraints and will impact to some degree where exactly matter dominance emerges from the radiative Universe.  We presume  that neutrino kinetic freeze-out completed before the onset of $e^+e^-$-annihilation into  photons. This means that  all entropy flows from $e^+e^-$-annihilation into the photons and leads to the ambient temperature of neutrinos in the current epoch being lower by a factor $(4/11)^{1/3}$. This is a common simplifying assumption, but small refinements are possible.  The current accepted model of the neutrino freeze-out is given in \cite{Mangano:2005cc}.

\begin{figure}
\begin{minipage}{\linewidth}
\makebox[0.5\linewidth]%
{\includegraphics[keepaspectratio=true,scale=0.55]{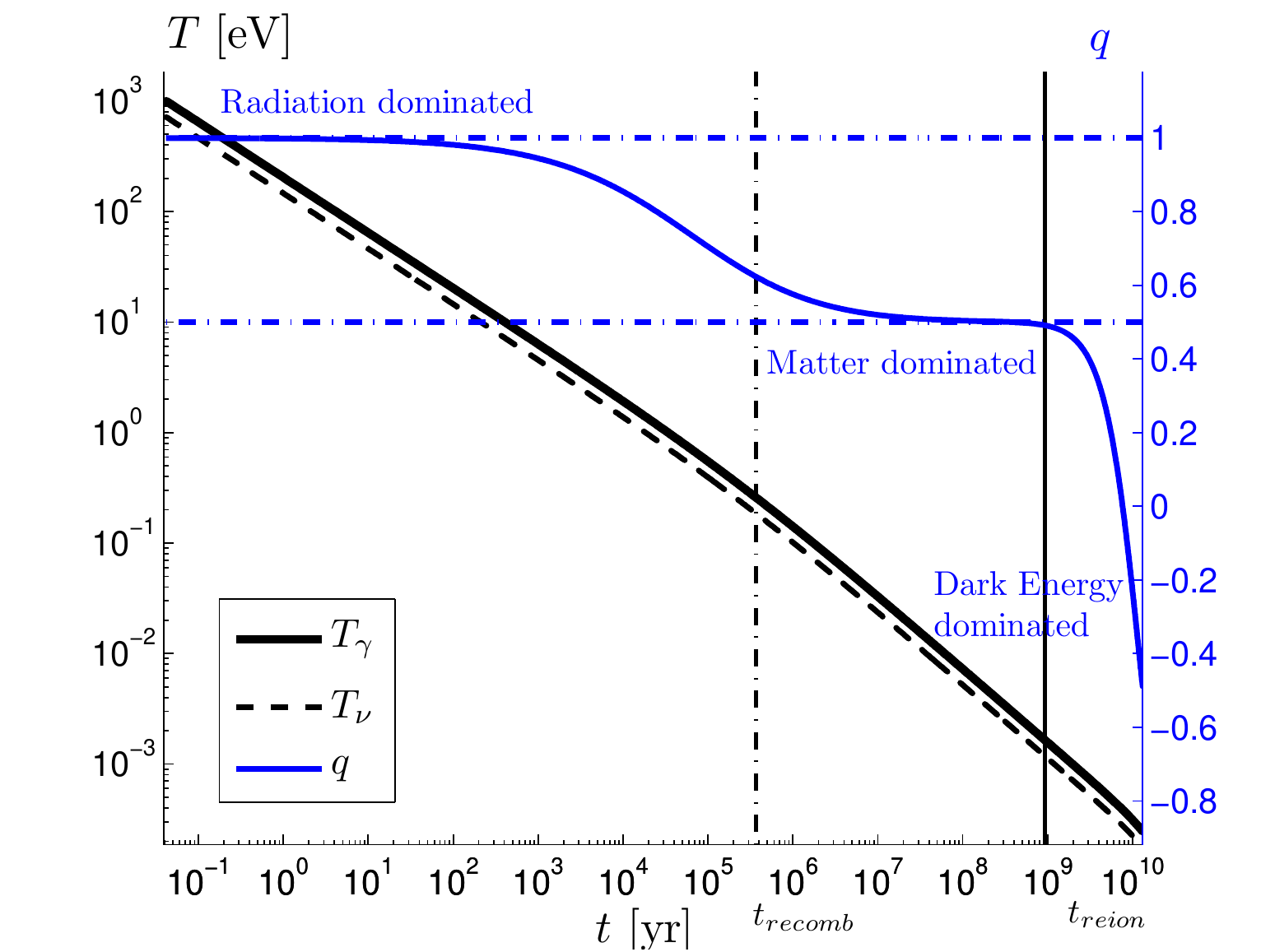}}
\makebox[0.5\linewidth]%
{\includegraphics[keepaspectratio=true,scale=0.55]{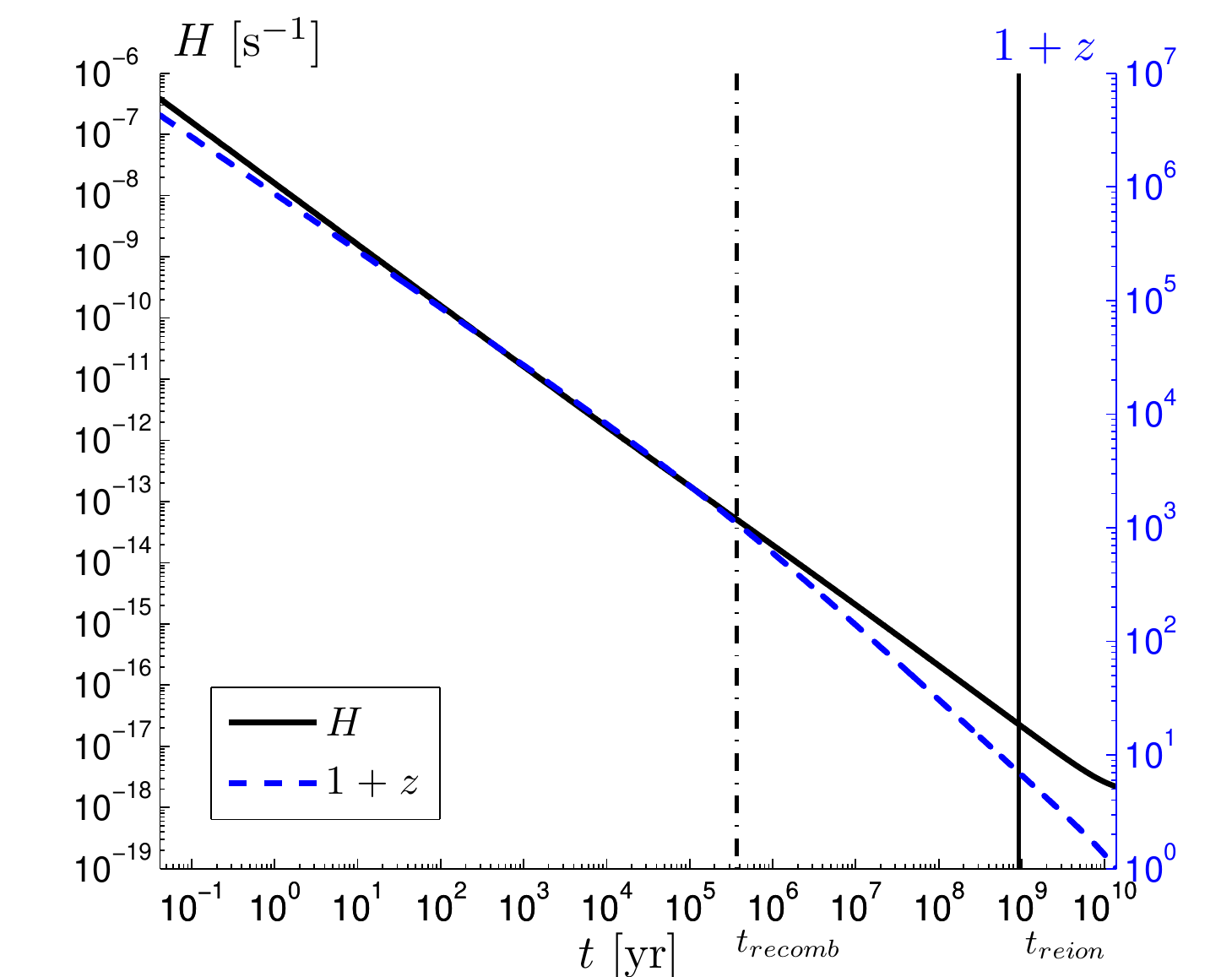}}
\caption{From  the present day  until  near BBN: on left -- evolution of temperature $T$  and deceleration parameter $q$; on right --  evolution of the Hubble parameter $H$ and redshift $z$.
\label{fig:today} }
\end{minipage}
\end{figure}

Figure \ref{fig:today} shows in the left frame the temperature  (left axis) and deceleration parameter (right axis)  from shortly after the completion of BBN  until today. This type of presentation will be repeated below for the two earlier eras; the connection from BBN to QGP and from QGP to the electroweak (EW) transition.  The horizontal dot-dashed lines show  the pure radiation-dominated value of $q=1$ and the matter-dominated value of $q=1/2$. The expansion in this era starts off as radiation-dominated, but transitions to matter-dominated starting around $T=\mathcal{O}(10\eV)$ and begins to transition to a dark energy dominated era at $T=\mathcal{O}(1\meV)$. We are still in the midst of this transition today. The dashed line shows the neutrino temperature, which maintains a constant ratio of $T_\nu/T_\gamma\approx (4/11)^{1/3}$ since neutrino decoupling and $e^+e^-$ annihilation around $T={\cal O}(1\MeV)$.  

The vertical dot-dashed lines show  the time of recombination at $T\simeq0.3\eV$ and   reionization at $T\simeq {\cal O}(1\meV)$. On the right in figure  \ref{fig:today}  we show the Hubble parameter $H$ and redshift $z+1\equiv a_0/a(t)$. We can see in figure \ref{fig:today} a visible deviation from power law behavior due to the transitions from radiation to matter dominated and from matter to dark energy dominated expansion.  These transitions are accentuated and more easily visualized in the form of the deceleration parameter $q$. The time span covered by the figure  \ref{fig:today} is in essence the entire lifespan of the Universe. All that happens before is a blip in comparison, but of course on a logarithmic time scale, the two eras we study next are of comparable duration.

\section{From BBN  passing through QED-neutrino plasma to the hadron Universe }\label{nudecoup}
The era separating the photon-neutrino-matter-dark energy Universe we just described from the QGP and hadron gas Universe is quite complex in its evolution.   It begins when the number of $e^+e^-$-pairs has decayed to the same abundance as the number of baryons in the Universe at the temperature  $T=\mathcal{O}(10\keV)$. It reaches back to $T={\cal O}(30\MeV)$ where there is no baryon antimatter as yet -- no antiprotons, antineutrons, antilambdas. However there is  significant abundance of muons, mesons, strangeness. Seen from the QGP side, this era begins after all baryon antimatter formed in QGP hadronization   has annihilated. Pions and muons disappear in abundance as compared to residual nucleons at $T \simeq {\cal O}(5\MeV)$, and we pass through neutrino kinetic freeze-out at $T\equiv T_k\simeq{\cal O}(1\MeV)$ followed by  $e^\pm$ annihilation and photon reheating at $T={\cal O}(m_e)$, and big bang nucleosynthesis (BBN)   at $T={\cal O}(100\keV)$ which actually is still immersed in a rich $e^+e^-$-pair `bath'. 

\begin{figure}
\begin{minipage}{\linewidth}
\makebox[0.5\linewidth]%
{\includegraphics[keepaspectratio=true,scale=0.54]{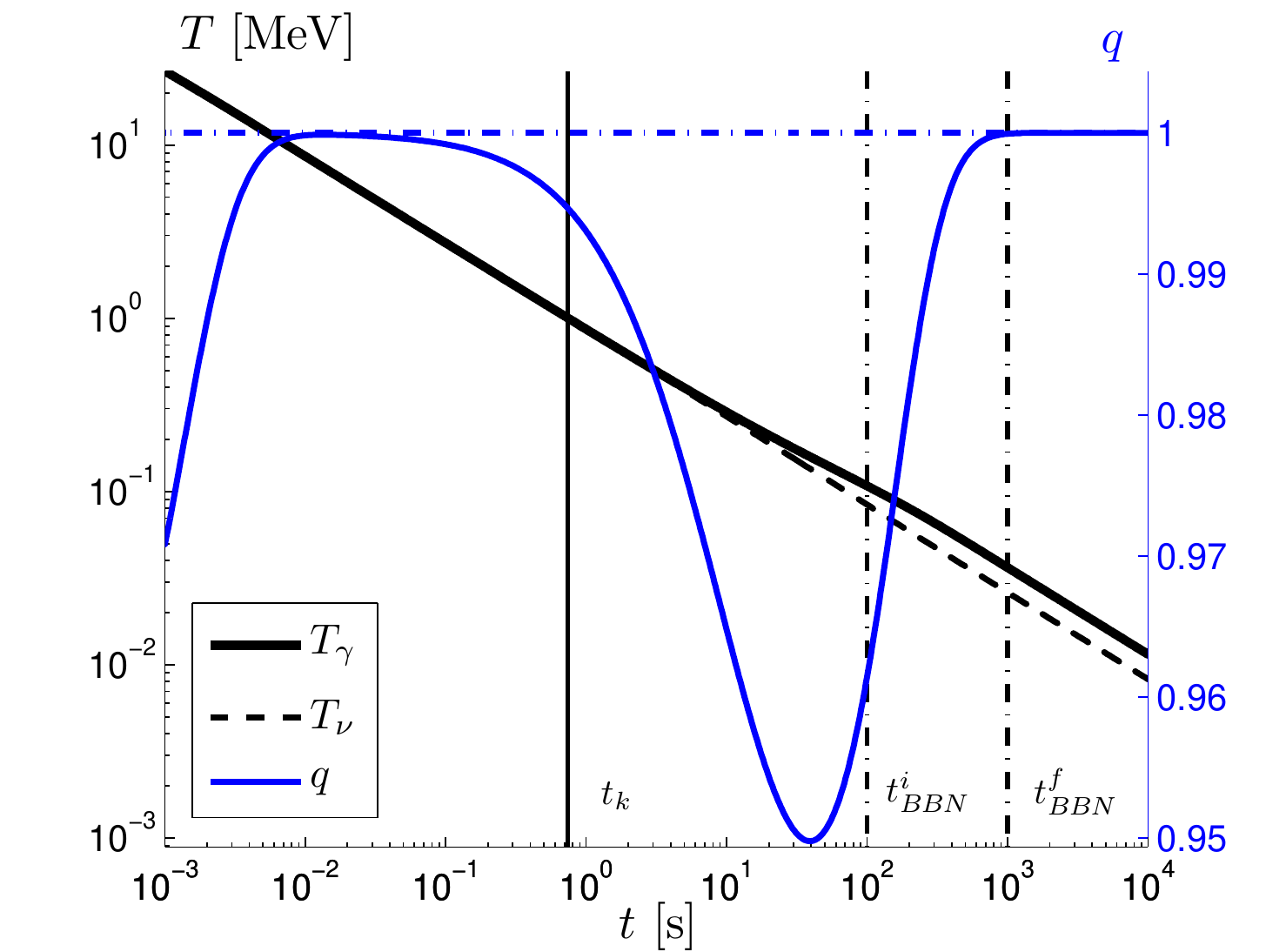}} 
\makebox[0.5\linewidth]%
{\includegraphics[keepaspectratio=true,scale=0.54]{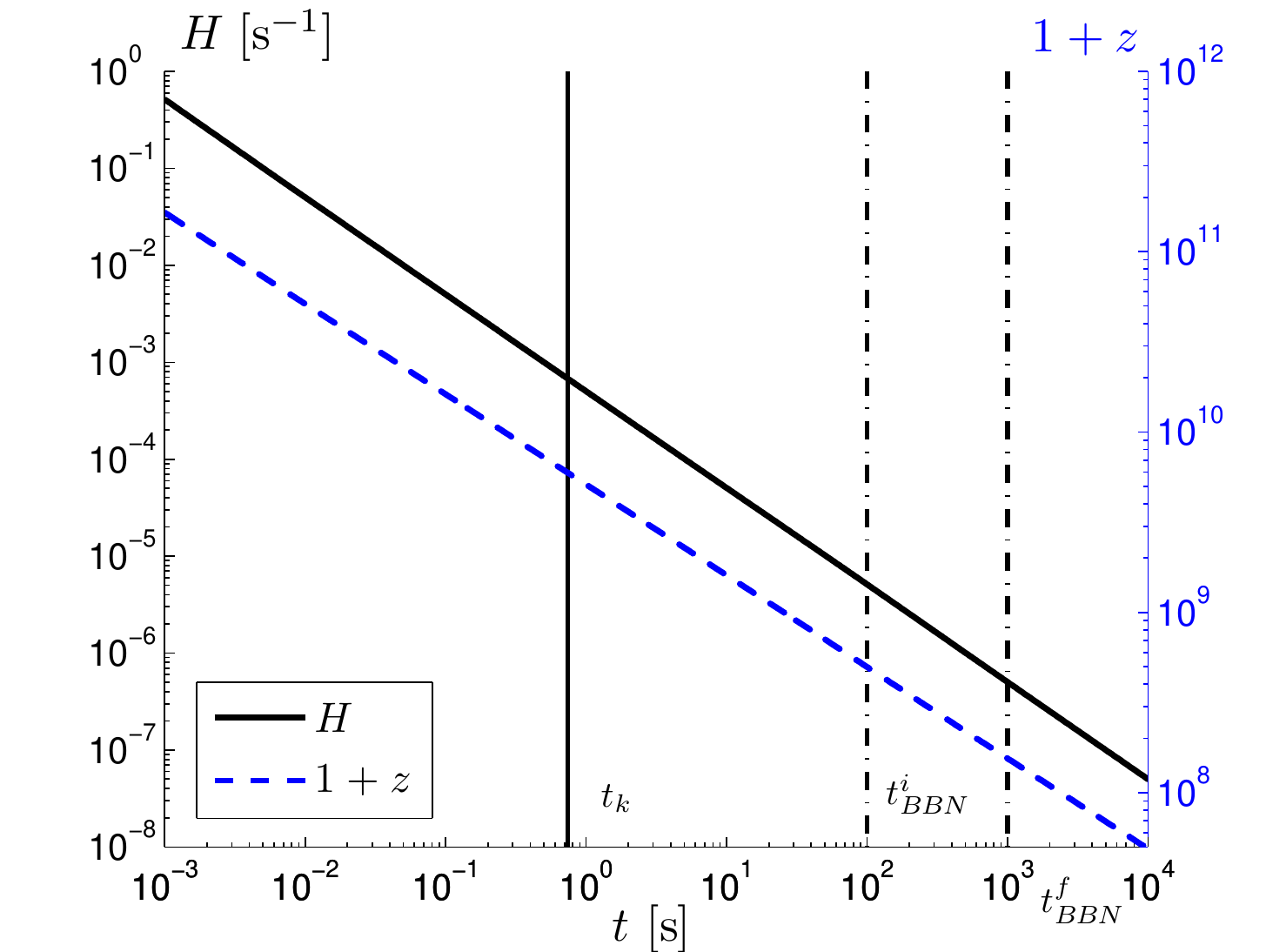}} 
\caption{From the end of baryon antimatter annihilation through BBN, see figure \ref{fig:today}.
\label{fig:BBN}  }
\end{minipage}
\end{figure}

 In figure~\ref{fig:BBN} the horizontal dot-dashed line for $q=1$  shows the pure radiation dominated value with two exceptions: the presence of massive pions  and muons reduce  the value of $q$ near to the maximal temperature shown, and when temperature is near the value of the electron mass the $e^+e^-$-pairs are not yet fully depleted but already sufficiently non-relativistic to cause another dip in $q$. The dashed line shows the neutrino temperature, which decouples from the $e^\pm$ and photon temperature at $T={\cal O}(1\MeV)$ when neutrinos freeze-out and begin free streaming. In figure~\ref{fig:BBN} the time unit is seconds, and the range spans the domain from fractions of a millisecond to a few hours. We left some time gap between this and the domain shown in figure \ref{fig:today}  describing the current era -- there is an uneventful evolution between the two domains.
 
\section{Connecting with the Quark Universe} \label{QGPera}
Going further back in time, the integration through the baryon antimatter  requires a detailed study of the hadron chemistry  using the methods originally developed for heavy ion collision physics as presented in Refs.~\cite{Fromerth:2012fe,Fromerth:2002wb}. There are some structural uncertainties in doing this computation which were discussed in these references. One should further note that the Universe was for a short time  filled with mesons, and most remarkably, a great number of diverse baryonic pairs.  At a time   of 10--15$\mu$s hadrons dissolve in the QGP phase. Freely propagating quarks or gluons filled the early Universe -- quarks remain unbound for as long as the Universe is above the critical temperature  $T_c\simeq 155\pm 10$\,MeV. The Universe is pushed apart by the thermal pressure  generated predominantly by  relativistic particles and thanks to the presence of numerous gluons which have additional degeneracy 8 from color charge compared to photons, and essentially massless $u,d,s$ quarks, Universe expansion was rather rapid.  

\begin{figure}
\begin{minipage}{\linewidth}
\makebox[0.5\linewidth]%
{\includegraphics[keepaspectratio=true,scale=0.54]{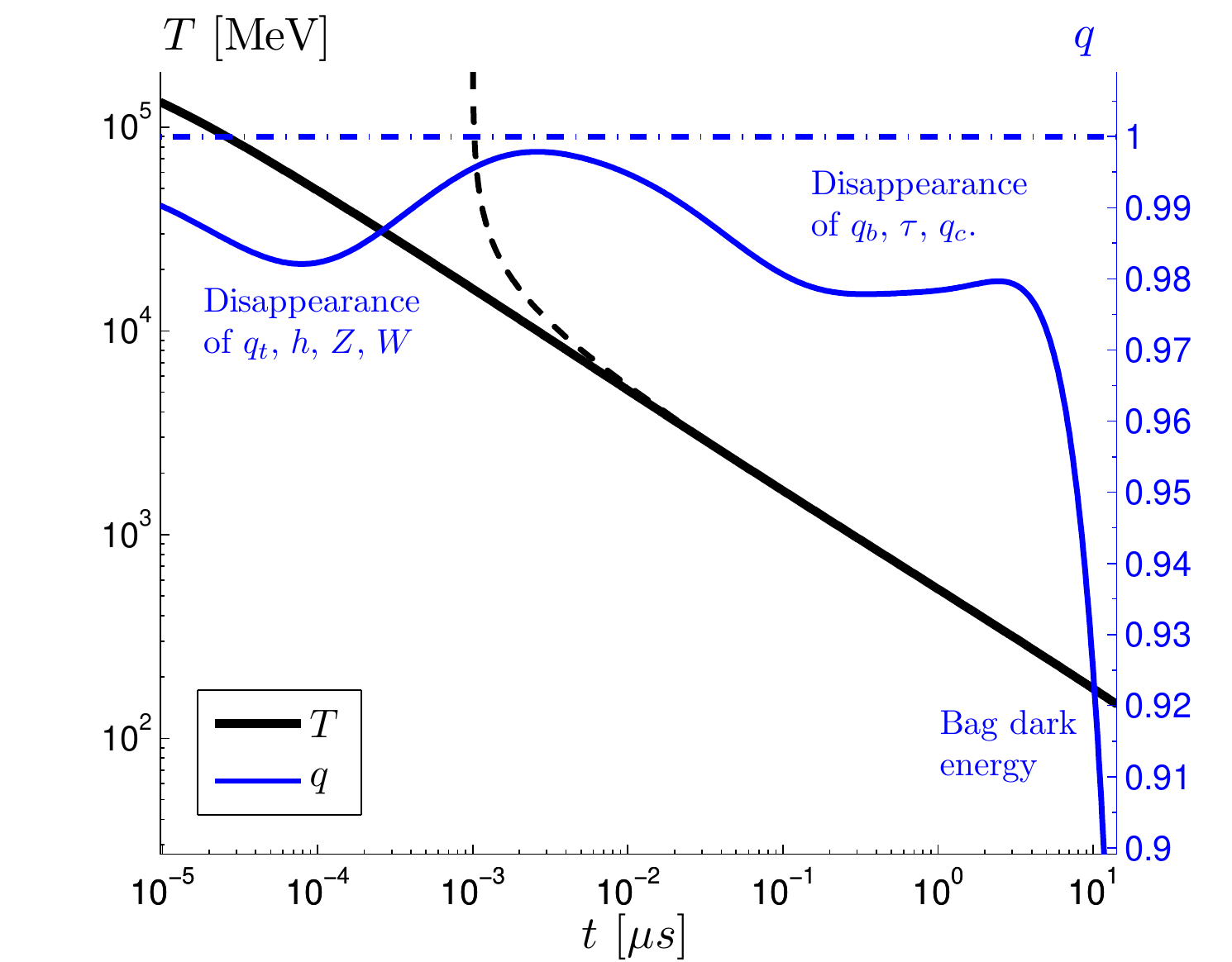}} 
\makebox[0.5\linewidth]%
{\includegraphics[keepaspectratio=true,scale=0.54]{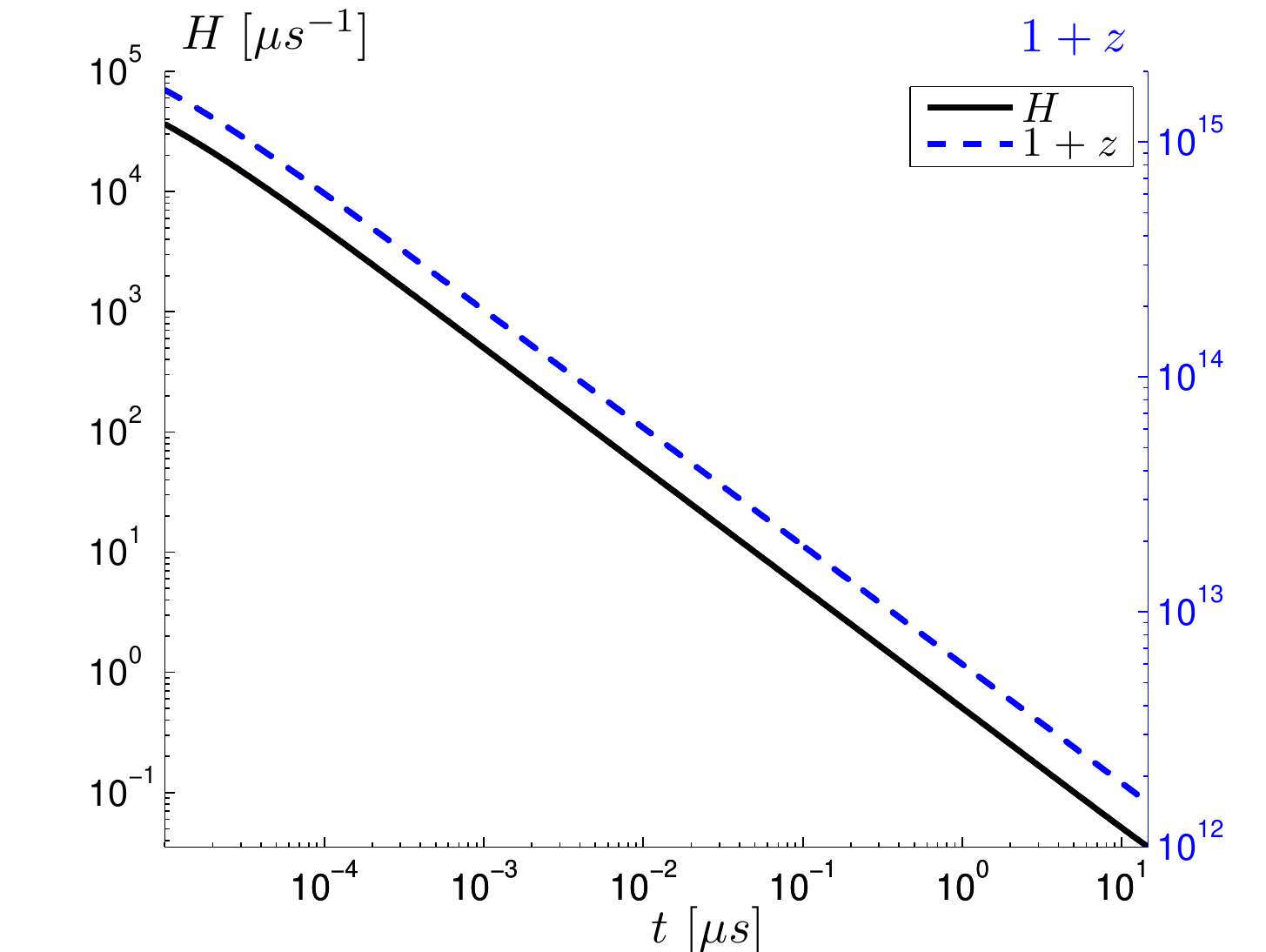}} 
\caption{From before our particle world to end of quark-gluon plasma, see figure \ref{fig:today}.
\label{fig:QGP}  }
\end{minipage}
\end{figure}

Figure~\ref{fig:QGP} shows the temperature  (left axis) and deceleration parameter (right axis)  from the end of the electroweak era at $T=\mathcal{O}(100\GeV)$, when the electroweak symmetry was broken and particles became massive,  through the quark gluon plasma (QGP) era and up to the QGP phase transition at $T={\cal O}(150\MeV)$. The horizontal dot-dashed line shows the pure radiation dominated value of $q=1$.  From the right hand scale, one can see that the expansion in this era is almost purely radiation dominated.  There are two smaller deviations from $q=1$ that correspond to $T$ on the order of the top, Higgs, Z, and W masses in the first case, and the bottom, tau, and charm masses in the other.  The comparatively larger drop in $q$ as $T$ approaches the QGP phase transition temperature arises because we injected the bag constant dark energy to demonstrate how a first order phase transition would accelerate the Universe dynamics.    On the right in figure~\ref{fig:QGP} we show the Hubble parameter and redshift through the QGP phase. Because the Universe is very nearly radiation dominated during this era, they obey simple power laws to high precision.  All nuances that can be seen in the deceleration parameter are lost at the double-logarithmic resolution  level.

The solution presented is the consequence of a forward integration of Einstein equations. A naive (i.e. ignoring any physics emerging at higher energy, including the details of the EW phase transition) backward integration establishes a pico-second timescale between the big bang and the end of the electroweak era. The precise timing is insignificant for the subsequent evolution and properties of the QGP era, as can be seen in the dashed line  in figure~\ref{fig:QGP}, where the starting time was shifted by $1$ns. On the other hand, this means that we can precisely show the time scale at which the presumed hadronization condition will be reached for a free quark-gluon gas. Inclusion of QCD interactions can be a future task.  

\begin{figure}
\begin{minipage}{\linewidth}
\makebox[\linewidth]{%
  \includegraphics[keepaspectratio=true,scale=0.6]{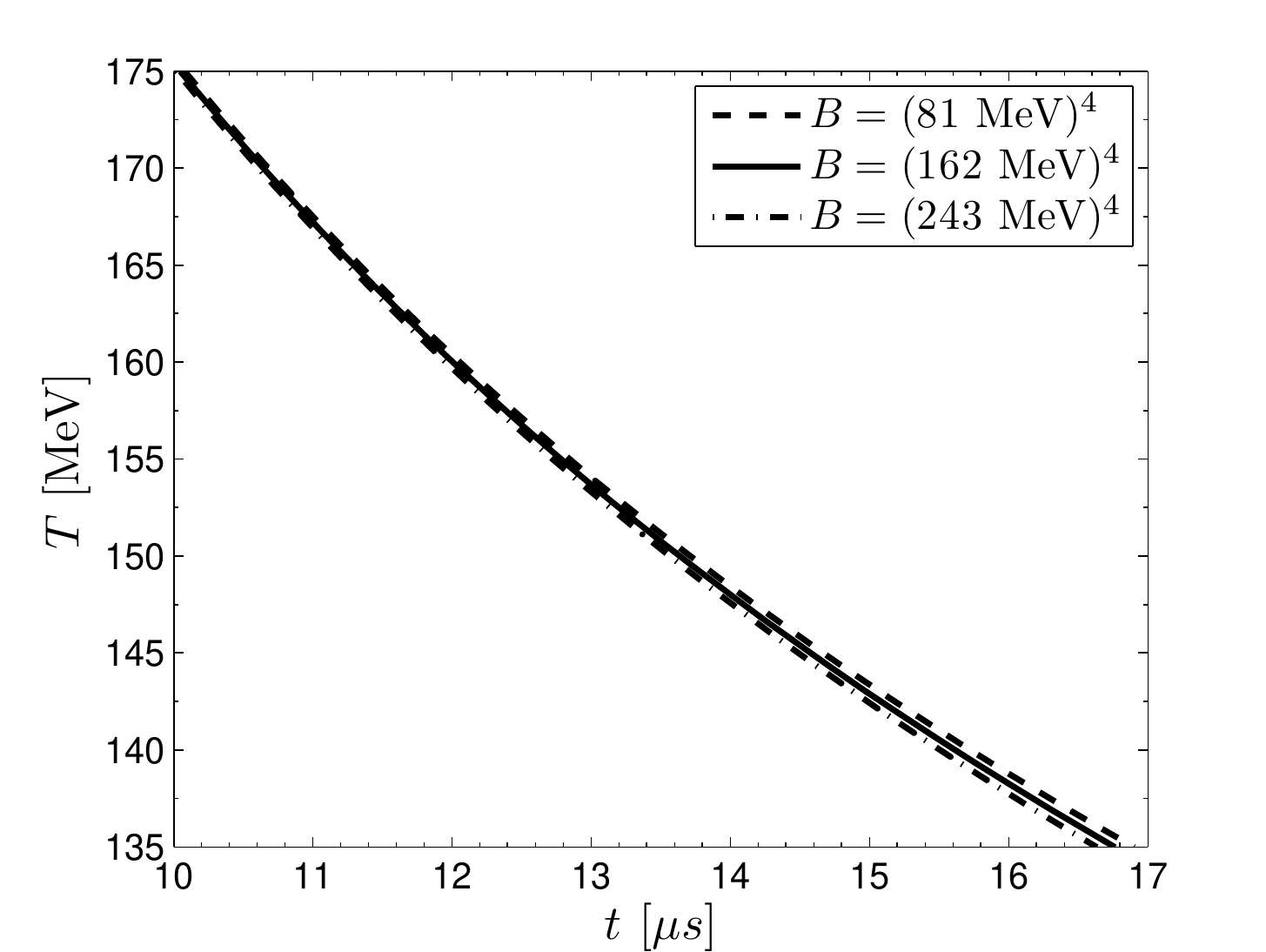}}
\caption{Evolution of temperature $T$ near the QGP phase transition for several values of the bag energy density.
\label{fig:QGP_zoom}}
\end{minipage}
\end{figure}

Figure~\ref{fig:QGP_zoom} zooms-in on the temperature range where the QGP phase transition occurs, with  several values of the bag energy density $\cal B$.  We see that the temperature dynamics are relatively insensitive to $\cal B$. On the other hand, $\cal B$ impacts strongly the critical temperature $T_c$ where the particle pressure cannot continue to support the deconfined state. This  determines the onset of the transition which will be determined entirely by the appropriate treatment of QCD degrees of freedom near the transition condition. Considering that quark Universe hadronization is expected at $T=155\pm10$ MeV within the QCD-lattice study of QGP\cite{Philipsen:2012nu}, we recognize in figure~\ref{fig:QGP_zoom} the probable life span  of 10--15$\mu$s  of the quark Universe.   The reason that we used a forward integration in the QGP era is that there are quite a few  uncertainties about the dynamics and physics of the slow hadronization process and we cannot yet fully characterize the time scales at which hadronization is complete as is discussed in in Refs.~\cite{Fromerth:2012fe,Fromerth:2002wb}. This also means that the redshift curve in figure \ref{fig:QGP} is only a (perhaps crude) approximation, as it will depend on the precise evolution of the scale factor $a(t)$ during hadronization. 

\section{Degrees of freedom and connection of all three eras of evolution}\label{Eralink}
Before  the QGP era, when the Higgs vacuum was not frozen, all particles were nearly massless, possibly retaining mass at the scale we observe today for neutrinos. Later  by the Higgs mechanism  mass was given to many of the QGP era particles. Before the Higgs vacuum froze,  the Universe was pushed apart by 28 bosonic and 90 fermionic degrees of freedom. Let us  verify: the doublet of charged Higgs particles has $4=2\times2=1+3$  degrees of freedom -- three will migrate to the longitudinal components of $W^\pm, Z$ when the electro-weak vacuum freezes and the EW symmetry breaking arises, while one is retained in the one single dynamical charge neutral Higgs component. In the massless stage, the SU(2)$\times$U(1) theory has 4$\times$2=8 gauge degrees of freedom where the first coefficient  is  the number of particles $(\gamma, Z, W^\pm)$ and each massless gauge boson has  two transverse polarizations. Adding in $8_c\times2_s=16$ gluonic degrees of freedom we obtain 4+8+16=28  bosonic degrees of freedom. 

The count of fermionic degrees of freedom includes three $f$ families, two spins $s$, another factor two for particle-antiparticle duality. We have in each family of flavors a doublet of $2\times 3_c$ quarks, 1-lepton and 1/2 neutrinos (due left-handedness which was not implemented counting spin). Thus we find that a total $3_f\times 2_p\times 2_s\times(2\times 3+1_l+1/2-\nu)=90$ fermionic degrees of freedom. We further recall that massless fermions contribute 7/8 of that of bosons in both pressure and energy density. Thus the total number of massless non-interacting particles at a temperature above the top quark mass scale, referring by convention to bosonic degrees of freedom, is is $g_{\rm SM}=28+90\times 7/8=106.75$ 

At times where dimensional scales are irrelevant and the expansion is radiative, entropy conservation means that  temperature scales inversely with the scale factor $a(t)$. This follows from \req{divTmn} when $ \varepsilon\simeq 3P   \propto T^4$. However, as the temperature drops and at their respective $m\simeq T$ scales, successively less massive particles annihilate and disappear from the thermal Universe. Their entropy reheats the other degrees of freedom and thus in the process the entropy originating in a massive degree of freedom is shifted into the other, still present  effectively massless degrees of freedom.  This causes the usual entropy conservation $T\propto 1/a(t)$ scaling to break down; during each of these `reorganization' periods the drop in temperature is slowed by the concentration of entropy in fewer degrees of freedom. To quantify this we study the ratio 
\beqn\label{redshiftratio}
r_U\equiv \left(\frac{1+z}{ T/T_{\rm now}}\right)^3, \qquad 1+z\equiv \frac{a_{\rm now}}{a(t)}.
\eeqn
The third power is chosen to connect with the growth of the volume of the Universe. 

\begin{figure} 
\centerline{\hspace*{0.4cm}\includegraphics[height=6.6cm]{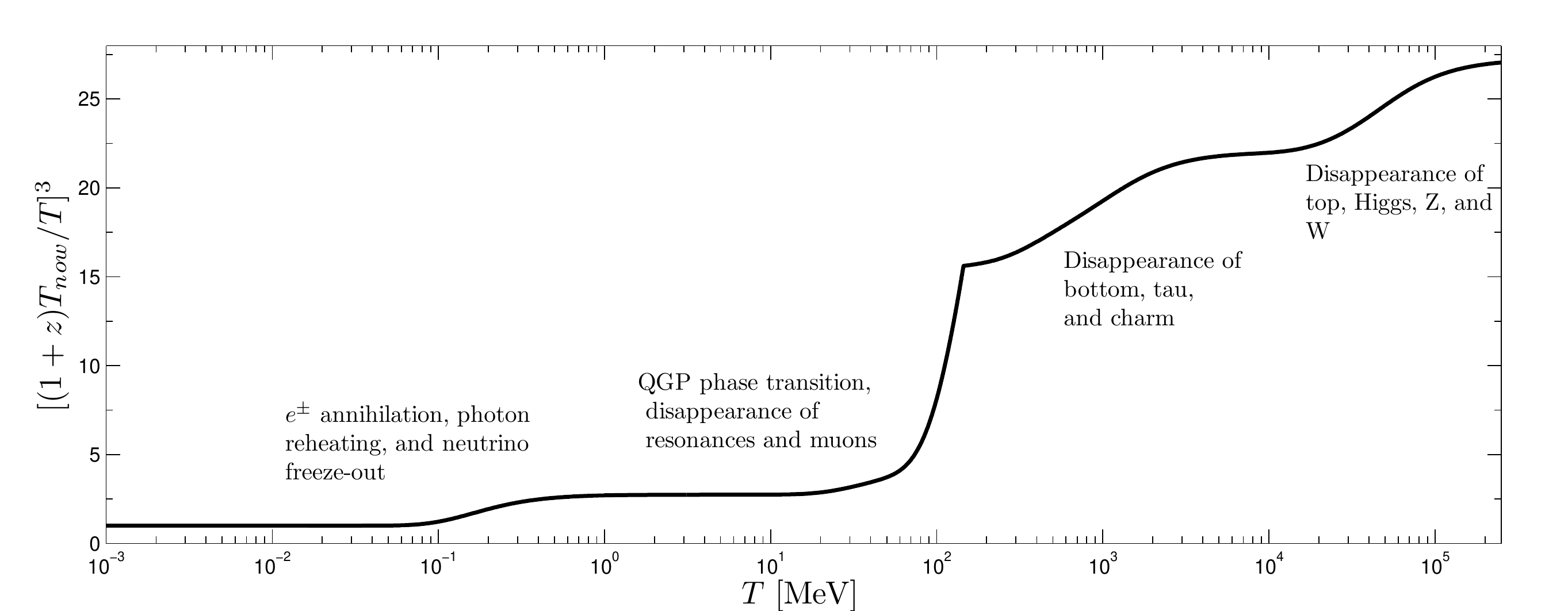}}
\caption{Disappearance of degrees of freedom through the evolution of the Universe in time and how this affects the fractional drop of temperature compared to red-shift. \label{fig:dof}}
 \end{figure}

In figure~\ref{fig:dof}  we show ratio $r_U$ \req{redshiftratio} as a function of photon temperature $T$ from the primordial high temperature $T$ in early Universe on right to the present on  the left, where $r_U$  must be by definition unity.  The periods of change seen in figure \ref{fig:dof} come when the temperature crosses the mass of a particle species that is in equilibrium. One can see drops corresponding to the disappearance of particles as indicated.   After $e^+e^-$ annihilation on left, there are no significant degrees of freedom remaining to annihilate and feed entropy into photons, and so $r_U$  remains constant until today. In producing this plot, we used a very rough model for the QGP-hadron phase transition. 

As long as the dynamics are at least approximately entropy conserving, the total drop in $r_U$ is entirely determined by entropy conservation. Namely, the magnitude of the drop in $r_U$ figure~\ref{fig:dof} is a measure of the inflation factor of the Universe, beyond what would be estimated from the thermal redshift.  We now show that it is also a measure of the number of degrees of freedom that have disappeared from the Universe. Consider   two times $t_1$ and $t_2$ at which all particle species that have not yet annihilated are effectively massless.  By conservation of comoving entropy and  scaling $T=1/a$ we have
\begin{equation}\label{r_ratio}
1=\frac{a_1^3S_{1}}{a_2^3 S_2}=\frac{a_1^3\sum_ig_i T_{1,i}^3}{a_2^3\sum_j g_j T_{2,j}^3},\qquad \frac{r_{U1}}{r_{U2}}=\frac{\sum_ig_i (T_{1,i}/T_{1,\gamma})^3}{\sum_j g_j (T_{2,j}/T_{2,\gamma})^3}
\end{equation}
where the sums are over the total number of degrees of freedom present at the indicated time and the degeneracy factors $g_i$ contain the $7/8$ factor for fermions. In the second form    we divided the numerator and denominator by $a_{now}T_{\gamma,now}$. We distinguish between the temperature of each particle species and our reference temperature, the photon temperature.  This is important since today neutrinos are colder than photons, due to  $e^+e^-$-reheating occurring after neutrinos decoupled. 

Using \req{r_ratio}  we  compute the total drop in $r_U$ shown in figure \ref{fig:dof}:  at $T=\mathcal{O}(100\GeV)$ the number of active degrees of freedom is slightly below $g_{\rm SM}=106.75$ due to the partial disappearance of top quarks, but this approximation will be good enough for our purposes.  At this time, all the species are in thermal equilibrium with photons and so $T_{1,i}/T_{1,\gamma}=1$ for all $i$.  Today we have $2$ photon and $7/8\times 6$ neutrino degrees of freedom with a neutrino to photon reheating ratio of approximately $T_\nu/T_\gamma=({4}/{11})^{1/3}$.  Therefore we have
\begin{equation}
\frac{r_{100GeV}}{r_{now}}= \frac{g_{SM}}{g_{\rm now}}=\frac{106.75}{2+\frac{7}{8}\times 6\times \frac{4}{11}}=27
\end{equation}
which is the  fractional change we see in figure \ref{fig:dof}. The meaning of this factor is that the Universe inflated by a factor 27 above the thermal red shift scale as massive particles disappeared successively from the inventory.

\section{Conclusions}\label{conclude}
Many open challenges of cosmic evolution deserving in depth  investigation come to mind. These are related to QGP-hadronization and matter emergence from baryon-antimatter annihilation~\cite{Fromerth:2012fe,Rafelski:2013qeu}, as well as neutrino transiting to a free-streaming decoupled gas~\cite{Birrell:2013gpa}. We list a few that we are either ready to address and/or we  already mentioned in this report: 
\begin{enumerate}  
\item The role of equilibrated flavor physics in  $u,d,s,c,b$ QGP and its hadronization; 
\item Degree of matter  inhomogeneity at QGP hadronization;
\item Strangeness is present in a significant abundance in early Universe down to $T=10$ MeV: potential for production of strange nuclearites;
\item What  happens during baryon antimatter annihilation, and how much entropy is produced;
\item How did the process of neutrino kinetic decoupling occur and in particular when precisely;
\item Are there effects in BBN due to the presence of dense $e^+e^-$-plasma.
\end{enumerate} 

We conclude by recalling that the baryon-matter asymmetry is an enigmatic riddle~\cite{Canetti:2012zc} of cosmology, of particle physics, and arguably, could be part of the physics of quark-gluon plasma and its hadronization transformation into matter. The study of QGP touches on other foundational challenges such as how  transport properties of the structured quantum vacuum, depending on temperature, allow quarks to move freely in the Universe, while when this motion freezes, the mass of particles is created.

\section*{\label{sec:acknowledgements}Acknowledgments}
This work has been supported by a grant from the U.S. Department of Energy, grant DE-FG02-04ER41318. JB is supported by the Department of Defense (DoD) through the National Defense Science \& Engineering Graduate Fellowship (NDSEG) Program.

\end{document}